# Controlling the Functional Properties of Oligothiophene Crystalline Nano/Micro-Fibers *via* Tailoring of the Self-Assembling Molecular Precursors

Francesca Di Maria,[§‡*] Mattia Zangoli,[§] Massimo Gazzano,[§] Eduardo Fabiano,[$] Denis Gentili,[†] Alberto Zanelli,[§] Andrea Fermi,[¥] Giacomo Bergamini,[Ø] Davide Bonifazi,[¥] Andrea Perinot,[#] Mario Caironi,[#] Raffaello Mazzaro,[±] Vittorio Morandi,[±] Giuseppe Gigli,[Δ,‡] Andrea Liscio,[¶§] Giovanna Barbarella[§ζ*]

§ *Istituto per la Sintesi Organica e Fotoreattivita' (ISOF),* ¶*Istituto per lo studio dei Materiali Nanostrutturati (ISMN),* ± *Istituto per la Microelettronica e i Microsistemi (IMM) and* ζ*Mediteknology srl, Consiglio Nazionale delle Ricerche, Via Gobetti 101, I-40129 Bologna, Italy*

‡ *Dpt. of Mathematics and Physics Ennio De Giorgi University of Salento, Lecce I-73100, Italy*

$ *Institute for Microelectronics and Microsystems (CNR-IMM), Via Monteroni, Campus Unisalento, 73100 Lecce, Italy and Center for Biomolecular Nanotechnologies @UNILE, Istituto Italiano di Tecnologia, Via Barsanti, I-73010 Arnesano, Italy*

† *CNR-ISMN, Via P. Gobetti 101, I-40129 Bologna, Italy*

¥ *School of Chemistry, Cardiff University, Park Place, CF10 3AT, Cardiff (UK)*

Ø *Dpt of Chemistry Giacomo Ciamician, University of Bologna, Via Selmi 2, I-40126 Bologna, Italy*

# *Center for Nano Science and Technology@PoliMi, Istituto Italiano di Tecnologia, Milano I-20133, Italy*

± *CNR-IMM, Via P. Gobetti 101, I-40129 Bologna, Italy*

Δ *CNR-NANOTEC Institute of Nanotechnology and Dept. of Mathematics and Physics Ennio De Giorgi University of Salento Lecce I-73100, Italy*

¶ *Istituto per la Microelettronica e i Microsistemi (IMM), Consiglio Nazionale delle Ricerche, Via del Fosso del Cavaliere 100, I-00133 Roma, Italy*

**ABSTRACT.** Oligothiophenes are π-conjugated semiconducting and fluorescent molecules whose self-assembly properties are widely investigated for application in organic electronics, optoelectronics, biophotonics and sensing. We report here an approach to the preparation of crystalline oligothiophene nano/micro-fibers based on the use of a 'sulfur overrich' quaterthiophene building block, -T4S4-, containing in its covalent network all the information needed to promote the directional, π–π stacking driven, self-assembly of Ar-T4S4-Ar oligomers into fibers with hierarchical supramolecular arrangement from nano- to microscale. We show that when Ar varies from unsubstituted thiophene to thiophene substituted with electron withdrawing groups, a wide redistribution of the molecular electronic charge takes place without substantially affecting the aggregation modalities of the oligomer. In this way a structurally comparable series of fibers is obtained having progressively varying optical properties, redox potentials, photoconductivity and type of prevailing charge carriers (from *p*- to *n*-type). A thorough characterization of the fibers based on SEM, CD, CV, X-ray diffraction, UV-vis and PL spectroscopies, photoconductivity and KPFM measurements is reported. With the aid of DFT calculations, combined with X-ray data, a model accounting for the growth of the fibers from molecular to nano- and microscale is proposed. We believe that the simple strategy outlined in this study allows to establish a straightforward correlation between the molecular structure of the components and the functional properties of the corresponding self-assembled nano/micro-fibers.

## INTRODUCTION

Investigations aimed to the design and synthesis of molecules capable to self-assemble into functional supramolecular structures by means of multiple non covalent interactions are at the forefront of nanoscience and nanotechnology.[1-10] Researches concern both equilibrium and dynamic, i.e. out-of-equilibrium, self-assembling systems, the latter being of outmost importance to understand the biological world.[1b,c] Since the properties of the supramolecular structures do not necessarily correspond to those of the single components, a crucial aspect of the investigations on molecular self-assembly is the search for building blocks storing sufficient information in their covalent framework to promote spontaneous organization into well defined supramolecular structures *via* precise non-bonded interactions (*specific non covalent interactional algorithms* according to J.M. Lehn's definition[11]). The identification of such molecular building blocks would enable the easy planning of supramolecular structures with programmed properties and function. Among supramolecular systems of major current interest are organic semiconducting fibers hierarchically organised across different length scales for application in photovoltaics, supramolecular electronics and nanophotonics.[2-7] Several organic self-assembled supramolecular functional fibers have been synthesized so far using various chemical strategies, including the preparation of nanochains made of organic nanoparticles.[3] Most of these self-assembly studies concern *p*-type semiconducting components (hole charge carriers) while *n*-type semiconducting components (electron carriers) have been much less explored. Recently, perylene imide molecules have been investigated for the formation of supramolecular *n*-type fibers and the relationship between molecular structure, noncovalent interactions, self-assembly conditions and fibers' morphology discussed in view of applications in optoelectronics.[6] Nevertheless further progress is required in design, synthesis, structural analysis and optical and electrical characterization of functional supramolecular fibers, while work-up simplicity, reproducibility and chemical stability are needed for any type of application. Tailoring the electronic properties of self-assembling supramolecular fibers while mantaining control over their crystalline structure and morphological habit has not yet been achieved. Moreover, contrary to small molecules or oligomers and polymers obtained by means of covalent synthesis, no full control on HOMO/LUMO energy gaps, redox properties and optical features of the fibers has been realized. Furthermore, the relevant properties of supramolecular fibers are generally optimized employing different molecular structures which self-organise in different manners making difficult to compare their performance. Among the components employed for the formation of supramolecular structures are thiophene derivatives.[12] Nanostructured crystalline fibers obtained by self-assembly of poly(3-hexylthiophene)[13] and of thiophene oligomers bearing terminal groups capable to form hydrogen bonds[14] have been reported. The formation of supramolecular fluorescent and *p*-type semiconducting fibers has also been reported by our group, by employing 'sulfur overrich' octithiophenes (having an extra sulfur per ring via β-functionalization with thioalkyl chains) as a strategy to promote anisotropic growth through directional S⋯S intra- and intermolecular non-bonded interactions.[15a,b] On these grounds it results clearly appealing to target a new class of oligothiophenes capable to self-assemble into functional supramolecular fibers and simultaneously to undergo wide electronic changes in charge distribution without substantially modifying the aggregation modalities. Such challenging perspective would enable unprecedented rational tuning of various properties within a set of comparable systems, in particular to pass from *p*-type to *n*-type charge transport characteristics. Here we report that 3,3',4'',4'''-tetrakis(hexylthio)-2,2':5',2'':5'',2''':5''',2''''-quaterthiophene (-**T4S4**-, Scheme 1) is a building block capable to promote the π−π stacking driven formation of functional crystalline fibers from **Ar-T4S4-Ar** oligothiophenes (compounds **1-5**) via solvent-vapour diffusion at room temperature.

**Chart 1.** Molecular structure of compounds **1-5**.

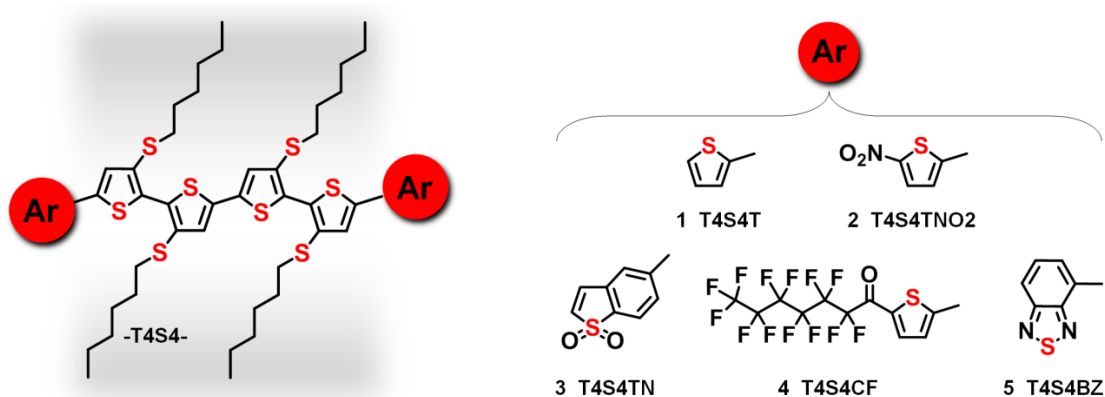

The addition of different terminal Ar groups to the -**T4S4**- building block allows to obtain a class of **Ar-T4S4-Ar** oligothiophenes whose electronic distribution is dictated by the **Ar** groups but that are nevertheless capable to form structurally similar fibers with hierarchical nano- to microscale supramolecular organization. The fibers are chemically and thermodynamically stable and grow with the same modalities on different substrates such as glass, gold, ITO, SiO$_2$. The variation of the electronic properties from **1** to **5** promotes the tuning of the functional properties of the fibers, in particular of their optical and electrical characteristics, with fluorescence spanning from visibile to near IR and prevailing charge carriers passing from *p*- to *n*-type. The properties and the morphology of the fibers are analyzed here by a variety of techniques: Scanning Electron Microscopy (SEM), Circular Dichroism (CD), Cyclic Voltammetry (CV), X-ray diffraction (XRD), UV-vis and PL spectroscopies, photoconductivity measurements and Kelvin Probe ForceMicroscopy (KPFM). Finally, with the aid of Density Functional Theory (DFT) calculations the characteristics of the intermolecular interactions of the self-assembling **1-5** are investigated. These are used, together with X-ray diffraction data, to propose a model accounting for the growth of the fibers from molecular to nano- and microscale.

**RESULTS AND DISCUSSION**

The head-to head regiochemistry of substitution and the alkyl chain length of the inner -**T4S4**-core in compounds **1-5** are crucial for promoting the self-assembly of **Ar-T4S4-Ar** oligothiophenes into fibers. We tested several other 'sulfur overrich' quaterthiophenes having different substitution patterns and different alkyl chain lengths and found them unable to promote the formation of supramolecular structures. In the Supporting Information we report the molecular structure, the synthesis and the characterization of all the tested quaterthiophenes, together with the synthesis and characterization of 3,3',3''',3'''',4'',4''''-hexakis(hexylthio)-2,2':5',2'':5'',2''':5''',2'''':5'''',2'''''- sexithiophene and of the the corresponding dodecamer also tested without any success.

*Synthesis of compounds* **1-5.** Compounds **1-5** were synthesized as described in Scheme 2. Compounds **1-5** were obtained in high yield starting from the monobromo (**1a**) or the monostannane (**1d**) of 3,3'-bis(hexylthio)-2,2'-bithiophene whose preparation is described in reference 15d. Compounds **1,3** and **5** were obtained from dibromoquaterhiophene **1c**, described in the same reference, via Stille coupling with the appropriate stannanes. Compounds **2** and **4** were obtained by Suzuki coupling thhrough the newly synthesized trimers **1e** and **1f** after bromination and subsequent treatment with bis(pinacolato) diboron with microwave assistance.

**Scheme 2.** Synthetic pattern for the preparation of compounds **1-5**.

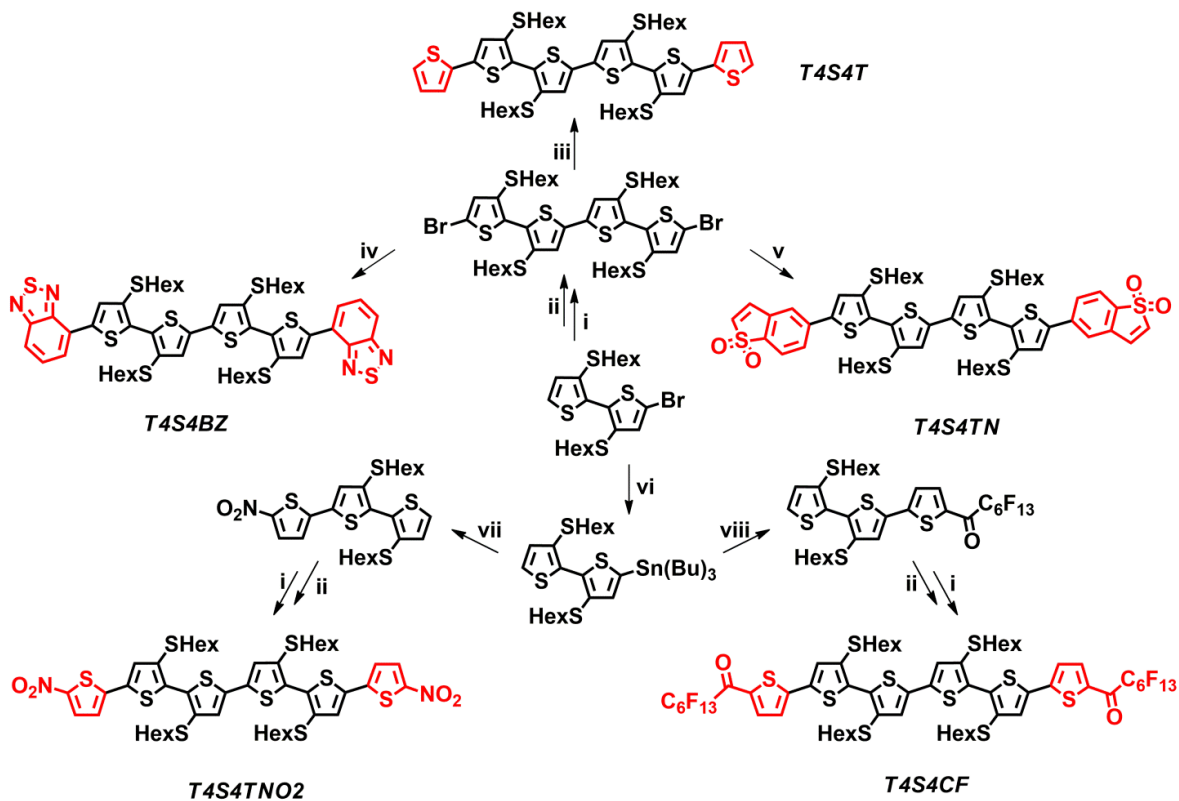

*i)* bis(pinacolato)diboron, NaHCO$_3$, Pd(dppf)Cl$_2$, MW, THF/H$_2$O, 80°C; *ii)* NBS, CH$_2$Cl$_2$, room T; *iii)* thiophene-2-ylboronic acid, NaHCO$_3$, Pd(dppf)Cl$_2$, MW, THF/H$_2$O, 80°C; *iv)* 4-(4,4,5,5-tetramethyl-1,3,2-dioxaborolan-2-yl)-2,1,3-benzothiadiazole, NaHCO$_3$, Pd(dppf)Cl$_2$, MW, THF/H$_2$O, 80°C; *v)* 5-(4,4,5,5-tetramethyl-1,3,2-dioxaborolan-2-yl)benzo[b]thiophene-1,1-dioxide, NaHCO$_3$, Pd(dppf)Cl$_2$, MW, THF/H$_2$O, 80°C; *vi)* n-BuLi, SnBu$_3$Cl, THF, -78°C; *vii)* 2-bromo-5-nitrothiophene, Pd(PPh$_3$)$_4$, toluene, 110°C; *viii)* 1-(5-bromothiophen-2-yl)-2,2,3,3,4,4,5,5,6,6,7,7,7-tridecafluoro-heptan-1-one, Pd(PPh$_3$)$_4$, toluene, 110°C.

For the synthesis of oligothiophenes via Stille and Suzuki coupling with the aid of ultrasound and microwave irradiation see reference 15f.

***Self-assembly of 1-5 into supramolecular fibers and their characterization.*** Compounds **1-5** spontaneously self-assemble into crystalline nanostructured fibers through solvent-exchange in solution, in which the molecules move from a good solvent (toluene) into a poor solvent (acetonitrile) and fast evaporation is inhibited.[16] Upon about one hour of exposition to acetonitrile vapours (see SI) of a 10$^{-3}$-10$^{-5}$ M solution of **1-5** in toluene, the formation of fibers with high aspect ratio and characterized by micrometric length and width and submicrometric thickness was observed (Figure 1). The formation of the fibers was reproducibile and afforded the same results independently of the type of substrate employed (glass, $SiO_2$, ITO, gold), indicating that thermodynamically stable supramolecular fibers are formed and that the morphologies are mainly determined by the molecular structure. All fibers were crystalline as shown by X-ray diffraction (see below) and optical microscopy. Figures S37, S38 and S39 display the optical microscopy images with cross polarizers of the fibers obtained from compounds **1-5**. Regardless of morphology they showed birefringence estinguishing in four positions with respect to the polarizers (every 90° rotation), indicating that the fibers had a high crystalline directional order. Figure 1 shows representative SEM images of the supramolecular fibers. It is seen that compounds **1**,**2** afford tape-like fibers whereas compounds **3**,**4** display the presence of helical superstructures at the nano and microscale and compound **5** forms polymorphic fibers, tape-like as well as helical. Despite the lack of stereogenic centers in the starting molecule, oligomers **3-5** form bundles of helices rolled into superhelical structures (coiled coils). Oligomer **3** (*T4S4STN*) affords fibers in the form of superhelices (helices of helices) and double helices of superhelices, all displaying the same-hand helicity (left-handed helicity, M type double helices). For this compound the inset in Figure 1 shows a large double helix of superhelices and indicates that the periodicity of the superhelices is size dependent. The SEM images relative to the fibers of compound **4** (*T4S4CF*) show the presence of nice superhelices of nanometer size (see the insets). Note that those shown in the inset of panel a display opposite handednesses. Concerning compound **5** (*T4S4BZ*), which forms polymorphic fibers with different morphology, we were able to obtain the tape-like fibers free from the helical ones (Figure 1, panel a) but not the latter in isolated form. Indeed, the helical fibers always grow on top of tape-like ones in the form of micrometer sized 'fusilli' (Figure 1, panel b) or as nanometer size coils wrapped around a micrometer size tape-like fiber (Figure 1, panel c). All fibers − whether tape-like or helical − displayed circular dichroism signals (Figure 1) corresponding to the π−π∗ transition in the UV-vis spectrum of the starting oligomer. Signal intensities ranged from a few mdeg (θ) for **1**, **3**, **5** to 40 mdeg (θ) for **4** to 100 mdeg (θ) for **2**.

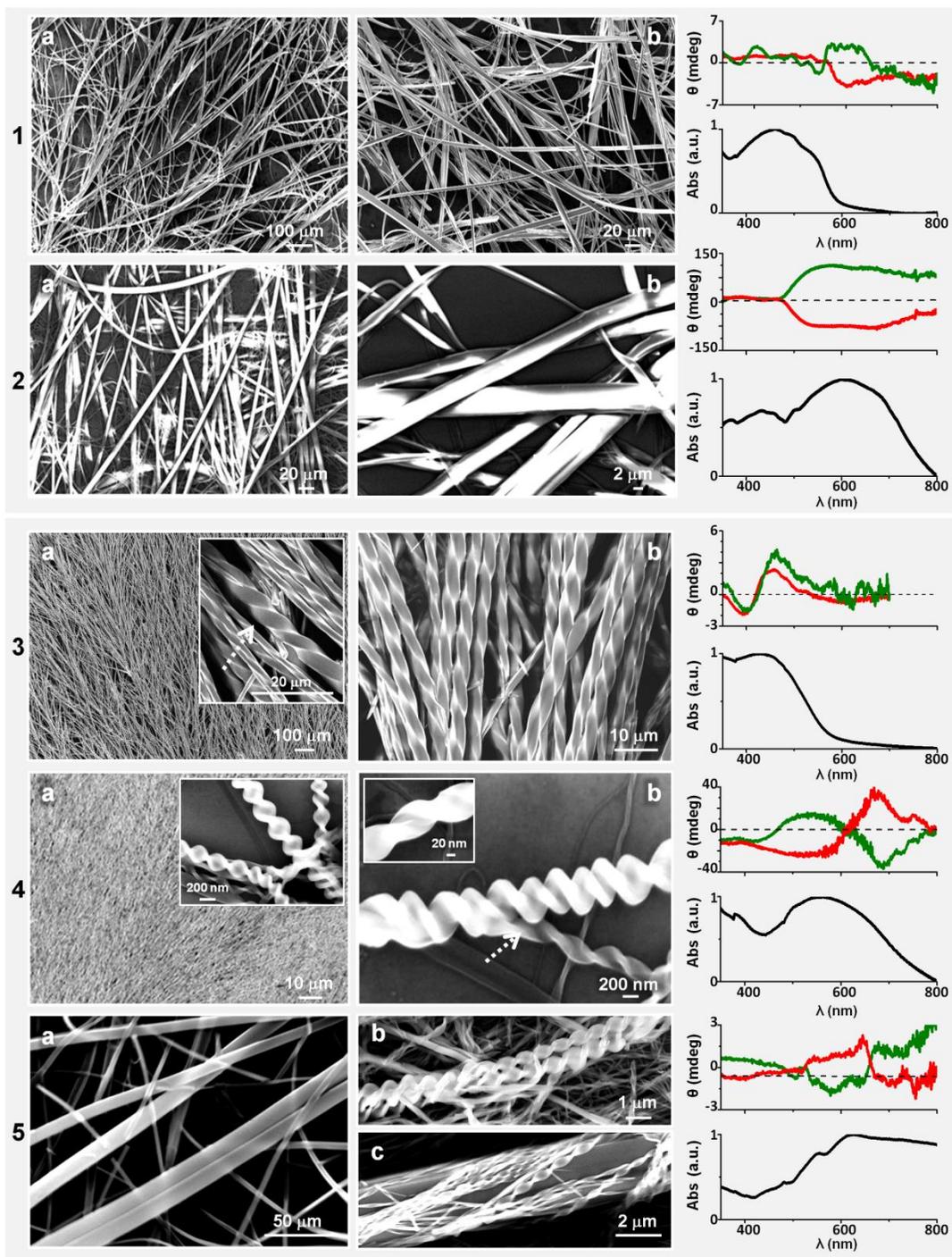

**Figure 1.** Scanning Electron Microscopy (SEM) images of fibers grown on glass from compounds **1**-**5** *via* the solvent-exchange method and Circular Dichroism plots of two different samples of the fibers of each compound.

Since **1-5** are centrosymmetric molecules and do not have chiral centers, no CD signals are observed in solution. Thus, the CD signals observed for the fibers are related to their supramolecular organization at the nanoscale. Indeed, it is well known that the main source of CD signals comes from adjacent interacting cromophores giving rise to excitons within a nanometer scale range.[17] One possible factor generating chirality in the centrosymmetric oligomers **1-5** are distorsions from planarity upon deposition of the molecules on the substrate. In this case, even if the bias of one or more thiophene rings from the molecular plane is very small, the molecules lose all symmetry elements and become chiral (conformational chirality). In addition (or in alternative), the interactions between two molecules − as soon as they come in contact, see the calculations below − also lead to distorsions hence to desymmetrization of the molecules. Subsequently, during the dynamic assembly process, chirality amplification[18] may take place generating an observable CD signal. Since upon deposition and/or intermolecular interaction the bias of a tiophene ring from the molecular plane (changes in interring angles ω) can occur in one or the opposite direction (+ω or −ω), both hand-helicities may be expected in different points of the same sample or in two different samples,[15a,b] resulting in CD signals displaying opposite Cotton effects. In agreement with expectations, the CD of all compounds but **3** display a Cotton effect of one sign and also of the opposite one corresponding to opposite helicity of the aggregates at the nanoscale. Nanoaggregates of opposite helicity are clearly visible in the SEM images of the fibers obtained from **4** (Figure 1, panel a) and **5** (Figure 1, panel c). The CD spectra of the tape-like fibers of compounds **1** and **2** reveal the presence of nanoscale aggregates of opposite helicity, which is lost at higher aggregation scales, possibly due to molecular packing effects. The fibers formed by **2** give rise to the largest CD intensity of all fibers. Since one of the factors determining the intensity of the CD signal is the proximity of the interacting cromophores,[17] this result is in agreement with the fact that the fibers formed by **2** are the most compact and of the highest crystalline quality (see X-ray diffraction section). Of all fibers only those pertaining to compound **3** afforded a CD signal displaying only one single (positive) Cotton effect. Despite repeated attempts on different samples and on different points of the same sample, we were unable to detect the CD signal of opposite sign. The simplest explanation for this behavior is that one of all possible conformations of **3** is largely preferred over all other ones but we do not have any experimental evidence for this. We limit ourself to note that the positive sign of the long wavelength part of the CD signal of **3** indicates the presence of a right handed helix at the nanoscale[17] while the micrometer sized double helices of superhelices observed in the SEM images are indicative of left-handed helicity (M type) at the microscale. Thus, the right-handed helices present at the nanoscale self-assemble into left-handed super helices at the microscale, a behaviour reminiscent of that of collagen[19] and also observed in other types of fibers.[20]

As soon as they are formed the self-assembled fibers of **1-5** are always randomly distributed on the different substrates used. However, the integration of the fibers into a device requires that the self-assembly takes place in a confined space dictated by the features of the substrate employed and that the characteristics of the fibers are mantained.[15d] In the present case we found that well-ordered aligned fibers can be obtained making use of the Lithographically Controlled Wetting (LCW)[21] technique to direct their self-assembly. Figure 2 shows well aligned fibers of compounds **2** (tape-like) and **4** (helical) grown with the LCW technique.

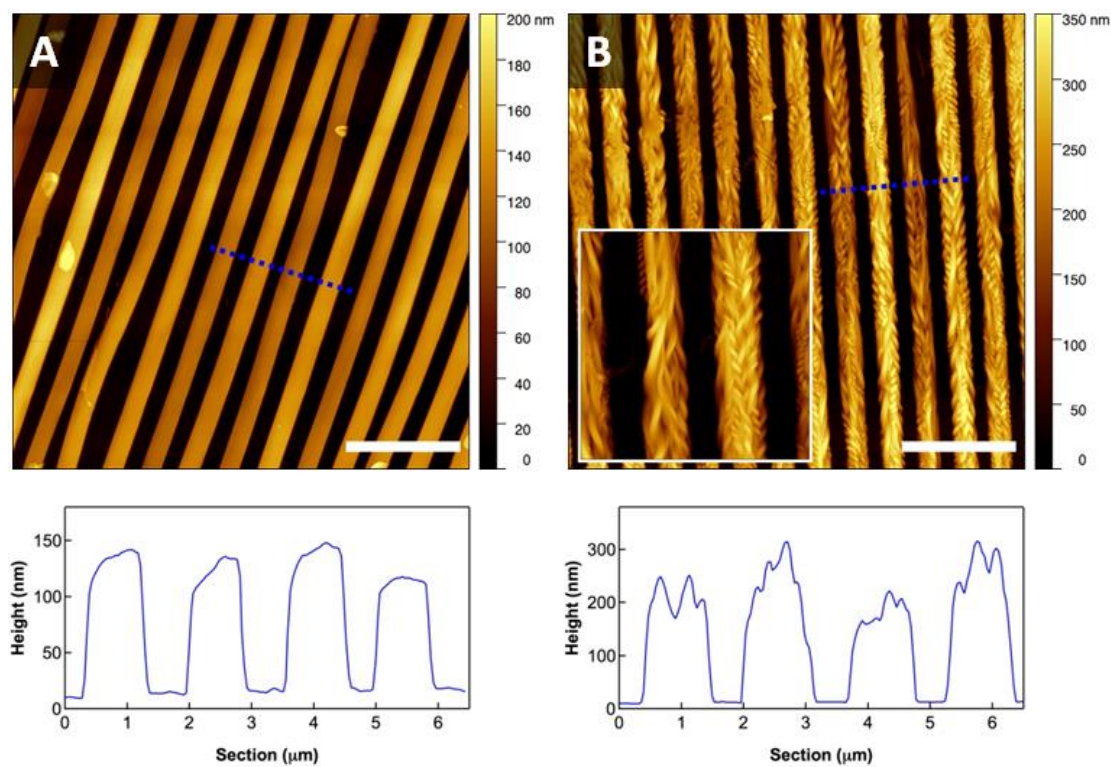

**Figure 2.** AFM images 20x20 µm$^2$ and related height plots of the fibers of: (A) **2** (***T4S4TNO2***) and (B) **4** (***T4S4CF***) patterned according the Lithographically Controlled Wetting (LCW)[21] technique. Scale bars = 5 µm. Inset image (B): zoom 5x5 µm$^2$.

A few drops of the compound dissolved in toluene were first placed on a silicon wafer, then a polydimethylsiloxane (PDMS) prepatterned stamp was placed on the top and the whole system exposed at room temperature (r.t.) for 24 h to the vapours of the (non-solvent) acetonitrile in a sealed container. In both cases parallel microfibers of similar dimensions, in particular width (1 µm) and height (150-300 nm) were obtained. The figure shows that while **2** affords straight tape-like fibers, **4** forms straight fibers made of tightly intertwined helical segments, indicating once again that even in a confined environment the morphology of the self-assembled fibers is mainly dependent on the molecular structure.

*Optical properties.* A comprehensive photo-physical analysis of compounds **1-5** in solution and as solid films was carried out. Table 1 reports the maxima in the absorption and emission spectra, the molar coefficients and the quantum yields of compounds **1-5** in solution, in cast films and as fibers while Table S1 reports the emission lifetimes in $CH_2Cl_2$ at r.t. The corresponding spectra are shown in Figure 3.

**Table 1.** Photophysical data of compounds **1-5.**

| *Item* | *Absorption* $\lambda$, nm ($\varepsilon$, $M^{-1} cm^{-1}$)[a] | *Emission* $\lambda_{max}$ (nm)[a], $\Phi$[a,b] | $\lambda_{max,DC}$ (nm)[c] | $\lambda_{max,fib.}$ (nm)[d], $\Phi$[d] |
|---|---|---|---|---|
| **1 T4S4T** | 421 (34700) | 538, 0.34 | 552 | 622, 0.006 |
| **2 T4S4TNO2** | 480 (41300) | 657, 0.004 | - | 840 |
| **3 T4S4TN** | 431 (38500) | 575, 0.20 | 626 | 649, 0.01 |
| **4 T4S4CF** | 482 (40200) | 667, 0.05 | 822 | 828 |
| **5 T4S4BZ** | 446 (30300) | 672, 0.03 | 674 | 687 |

[a]In $CH_2Cl_2$ at r.t. on air-equilibrated solutions. [b]Determined using $[Ru(bpy)_3]Cl_2$ in $H_2O$ ($\Phi$=0.04).[22-23] [c]Recorded from drop-cast amorphous thin films on glass surfaces. [d]Recorded from crystalline fibers grown on glass surfaces. For solid samples, emission quantum yield was calculated from corrected emission spectra registered by an Edinburgh FLS920 spectrofluoremeter equipped with a LabSphere barium sulfate coated integrating sphere (4 in.), a 450W Xe lamp (excitation wavelength tunable by a monochromator supplied with the instrument) as light source, and a R928 photomultiplayer tube, following the procedure described in reference 24.

All compounds show broad absorption bands associated to the lowest energy π-π* transition of the central thiophenic backbone, with the corresponding maxima shifting to higher wavelengths in the presence of strong electron-withdrawing groups such as -NO$_2$ and -COC$_6$F$_{13}$ as the outermost substituents. As a consequence, solutions of **1-5** cover a range of colors from yellow (**1**, *T4S4T*) to wine red (**2**, *T4S4TNO2*), with extinction coefficients included between *circa* 30000 and 41000 M$^{-1}$cm$^{-1}$.

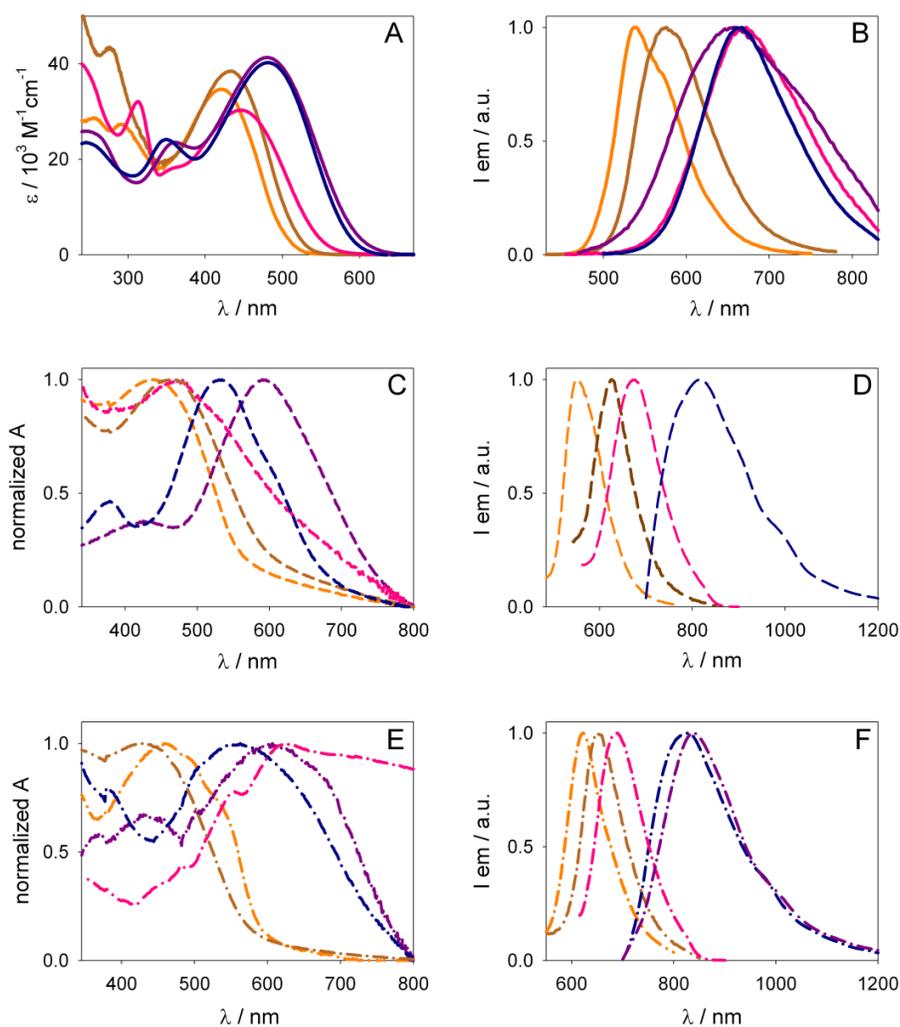

**Figure 3.** Absorption and emission spectra of **1** (*T4S4T*, yellow), **3** (*T4S4TN*, brown), **5** (*T4S4BZ*, pink), **4** (*T4S4CF*, blue) and **2** (*T4S4TNO2*, purple) in different conditions. A-B: air-equilibrated CH$_2$Cl$_2$ at r.t; C-D: drop-cast films on glass surfaces at r.t.; E-F: as crystalline fibers deposited on glass surfaces at r.t.

Emission spectra in $CH_2Cl_2$ at r.t. show a similar trend (Figure 3B), with maxima red-shifting in the visible spectrum in a range between 540 and 670 nm, while quantum efficiencies result strongly dependent on the type of substituent, dropping from 0.34 for **1** to 0.004 for **2**. With the exception of **2**, drop-cast samples from $CH_2Cl_2$ solutions showed appreciable luminescence signals, with **4** displaying a broad emission profile with a maximum in the NIR region around 820 nm (Figure 3D). As for the emission quantum yields of amorphous films of **1**-**5**, likely due to the increased predominance of packing-induced radiationless deactivation pathways[25-26], they are lower compared to those recorded in solution and below the limit of detection of our instrumentation ($\Phi$<0.005). Emission lifetimes (see Table S1), determined in the same conditions, point out singlet excited state deactivations with sub- nanosecond decays. For the majority of the molecules taken in exam, multiexponential decays were evidenced, possibly due to inhomogeneities between the adopted different conformations in solution and in amorphous films, as revealed for other thiophene-based oligo- and polymeric structures.[27-28] Moreover, in drop-cast films of **1**-**5**, the relatively strong interactions occurring intermolecularly likely affect their absorption profiles (Figure 3C), showing the broadening and the red-shift of the lowest energy absorption maxima with respect to the signals recorded in solution. This trend is even more evident when molecules organize in crystalline fibers on glass surfaces (Figure 3E), which display broader absorptions at lower energies (up to 800 nm for **2** and **4**) compared to those recorded on drop-cast films. In the case of **5** (*T4S4BZ*), the wide-ranging absorption spectrum of the crystalline fibers reflects the presence of different polymorphs in the same sample (see also Figure 1). More interestingly, deposited fibers show appreciable luminescence ($\Phi$~0.01 for **1** and **3**), ranging from the orange-red of **1** towards the NIR spectrum (Figure 4F). More specifically, **4** and **2** show their fluorescence maximum around 830 and 840 nm, respectively, with a broad emission profile extending up to 1200 nm (~1.05 eV).

*Cyclic Voltammetry.* Table 2 reports the redox potentials of compounds **1-5** in solution, cast films from $CH_2Cl_2$ and as fibers. The corresponding cyclovoltammetries are reported in Supporting Information (Figure S40) together with a few comments. In the present context the most significant result is that for all compounds the reduction potential of the fibers is significantly less negative than that measured in solution and in cast film, indicating that self-assembly into fibers leads to the increase in electron affinity. For example, compound **1** in solution and in cast film has a reduction wave outside the electrochemical window of the electrolyte (< -1.8 V) while the corresponding fibers show a reduction potential of -1.41 V, indicating an increases in electron affinity of about 400 mV. Also compounds **2-5** show less negative reduction potentials when grown in solid state with respect to the solution and, in particular, when the order of the solid is driven towards the fibers. The displacement of the reduction potentials towards less negative values in the fibers compared to the solution varies from 80 mV in **2** (*T4S4TNO2*) to 430 mV in **5** (*T4S4BZ*).

**Table 2.** Cyclic Voltammetry data of **1-5**.[a]

| Item | | $E_{ox}/V$ | $E_{red}/V$ | $E_g/eV$ |
|---|---|---|---|---|
| **1** *T4S4T* | Solution | 0.76 | <-1.8 | > 2.56 |
| | Cast film | 0.71 | <-1.8 | >2.51 |
| | **Fibers** | **0.69** | **-1.41** | **2.10** |
| **2** *T4S4TNO2* | Solution | 0.87 | -0.82 | 1.69 |
| | Cast film | 0.75 | -0.83 | 1.58 |
| | **Fibers** | **0.79** | **-0.74** | **1.53** |
| **3** *T4S4TN* | Solution | 0.81 | -1.42 | 2.23 |
| | Cast film | 0.66 | -1.39 | 2.05 |
| | **Fibers** | **0.71** | **-1.15** | **1.86** |
| **4** *T4S4CF* | Solution | 0.74 | -1.13 | 1.87 |
| | Cast film | 0.74 | -1.03 | 1.77 |
| | **Fibers** | **0.90** | **-0.78** | **1.68** |
| **5** *T4S4BZ* | Solution | 0.67 | -1.33 | 2.00 |
| | Cast film | 0.73 | -1.24 | 1.97 |
| | **Fibers** | **0.78** | **-0.90** | **1.68** |

[a] Solution: Vs SCE in $CH_2Cl_2$ 0.1 mol $L^{-1}$ $(C_4H_9)_4NClO_4$. Solid state: Vs SCE in propylene carbonate 0.1 mol $L^{-1}$ $(C_2H_5)_4NBF_4$.

Passing from the solution to the fibers the oxidation potential *decreases* by 70 mV in **1**, 80 mV in **2**, 100 mV in **3**. However, the opposite trend is observed in **4** and **5**, as the oxidation potential *increases* by 160 mV and 110 mV, respectively. As a result for all compounds there is a net *decrease* in the energy gap from solution to the fibers, the lowest energy gap observed being that of compound **2** (1.53 eV).

***DFT calculations.*** To understand the basic interactions governing the aggregation in fibers we studied the dimers of molecules **1-5** using density functional theory (DFT) calculations. Ground-state calculations and geometry optimizations were performed with the TURBOMOLE program package[29-30] using the BLOC exchange-correlation[31-33] functional and the multipole accelerated resolution of identity approximation.[34] For isolated molecules a def2-TZVP basis set[35] was employed. For the dimers, the optimal structures have been identified performing a preliminary screening, via geometry optimizations at the BLOC/def2-SV(P) level of theory,[31-33,36] on a large number (about 100) of guess structures generated by random displacements of two monomers. Subsequently, the 20 best candidates have been optimized at the BLOC/def2-TZVP level to identify the most stable configuration. The calculated interaction energies of the dimers [kcal/(mol*atom)] are the following: 0.13 **(1, T4S4T)**; 0.19 **(2,T4S4NO2)**; 0.17 **(3,T4S4TN)**; 0.12 **(4, T4S4CF)**; 0.15 **(5, T4S4BZ)**. Their structure is depicted in in the X-ray diffraction section below (see Figure 7). We have found that in all cases, partially excepting compound **4**, the π−π stacking between the molecular backbones plays a major role. This is very evident in the dimer of sexithiophene **1**, where the π−π stacking is basically the only force binding the two molecules. Adding side substituents or varying the terminal groups, however, additional effects start to be relevant. These are related to a significant charge redistribution in the molecules, which is visualized in Figure 4, where the electrostatic potential generated by each molecule are reported. In the case of **5**, having terminal benzothiadiazole groups, the net effect is not very large and the molecules in the dimer show only small distortions with respect to the isolated ones. This traces back to the fact that, despite an important charge depletion on the thiadiazole moiety, the charge distribution in the molecular backbone and the side hexyls is not much perturbed. Thus, the main bonding character is still a π−π stacking. On the contrary, in compounds **3** and **2**, the presence of oxygen atoms in the terminal substituents, induces a stronger polarization, with charge depletion in the hexyls (and partially in the backbone) and charge accumulation on the oxygens. As a consequence, when two molecules interact, hydrogen bonds can form. These involve the oxygen atoms and some hydrogens in the hexyls (or, to a less extent, in the backbone).

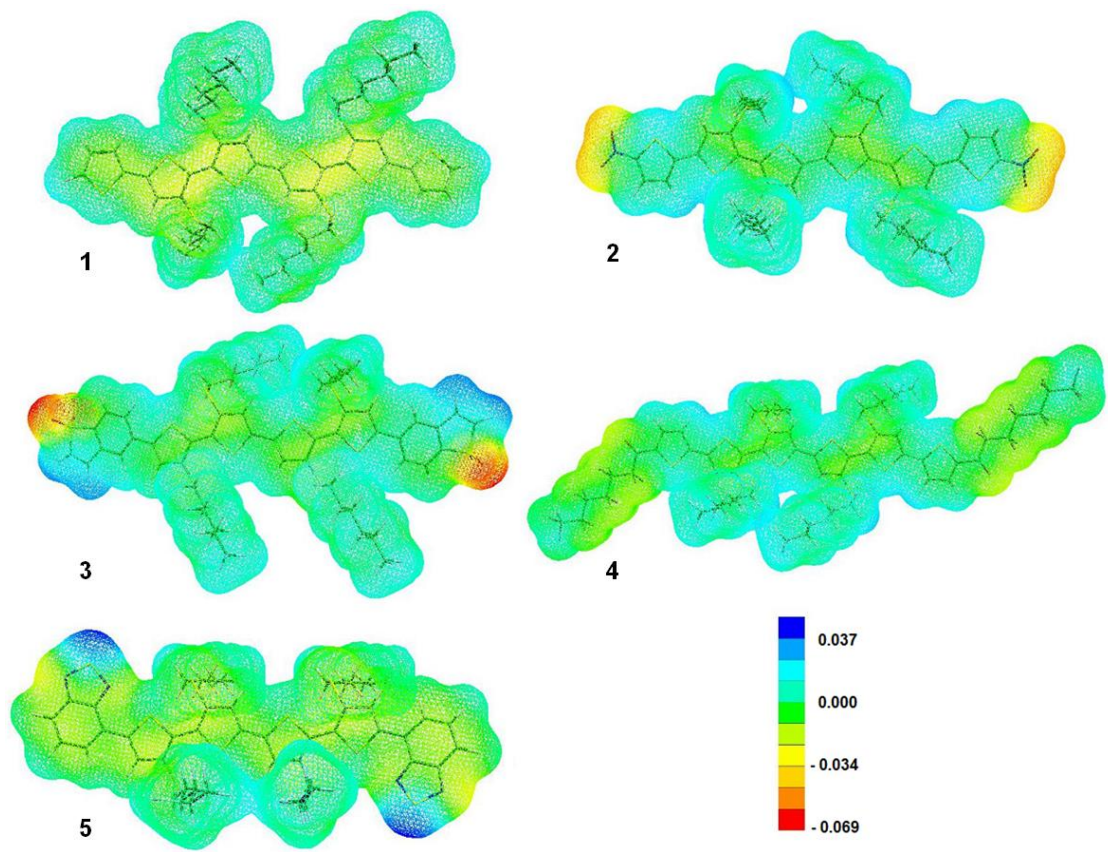

**Figure 4.** Electrostatic potential (in au) on the van der Waals surface for isolated **1-5**.

To this purpose, we have performed a study using the NCI indicator[37-38] − which can identify hydrogen bonding regions and highlight them in real space − using the NCI PLOT program.[39] The corresponding plot is reported in Figure 5, and shows indeed hydrogen-bond regions between the oxygen atoms and the hydrogens (or the backbone). Nevertheless, because of fine differences between the two molecules and especially due to the different position of the oxygens (out of plane in **3**, in plane in **2**), the final dimers' structures are different in the two cases. For **3**, we observe prominently hydrogen bond interactions involving the hexyls, thus in the dimer the two molecules tend to twist onto each other, despite the molecular backbone is not strongly distorted.

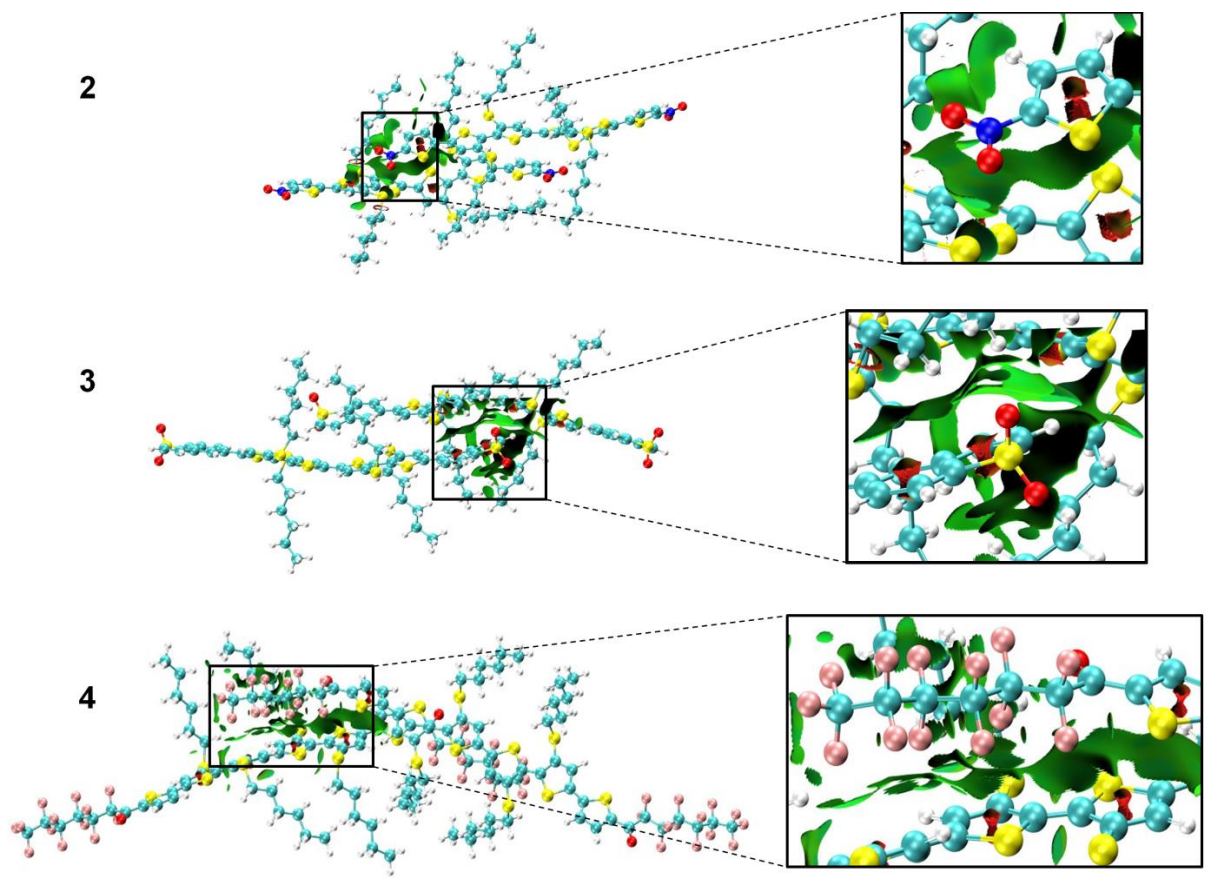

**Figure 5.** NCI indicator isosurface plots for the dimers of compounds **2** (***T4S4TNO2***), **3** (***T4S4TN***) and **4** (***T4S4CF***) identifying hydrogen bonding regions (the green regions between the two molecules). For clarity the isosurfaces were generated only for the regions indicated by the boxes.

On the other hand, in **2**, the oxygen atoms do not interact with the hydrogens in the hexyls but preferably with the molecular backbone. Consequently, instead of a twisting of the two molecules around each other, we rather observe strong distortion of the backbone of each individual molecule. Finally, considering compound **4** we see that the presence of many fluorine atoms induces a significant depletion of charge in the π system of the molecular backbone with consequent accumulation on the fluorines. Accordingly, in the dimer we do not observe anymore a significant π–π stacking interaction but rather an interaction between the fluorines of one molecule and the backbone of the other molecule.

***X-ray diffraction.*** All the fibers were analyzed by X-ray diffraction and found to be highly crystalline. The X-ray diffraction plots of the microfibers from **1-5** are shown in Figure 6.

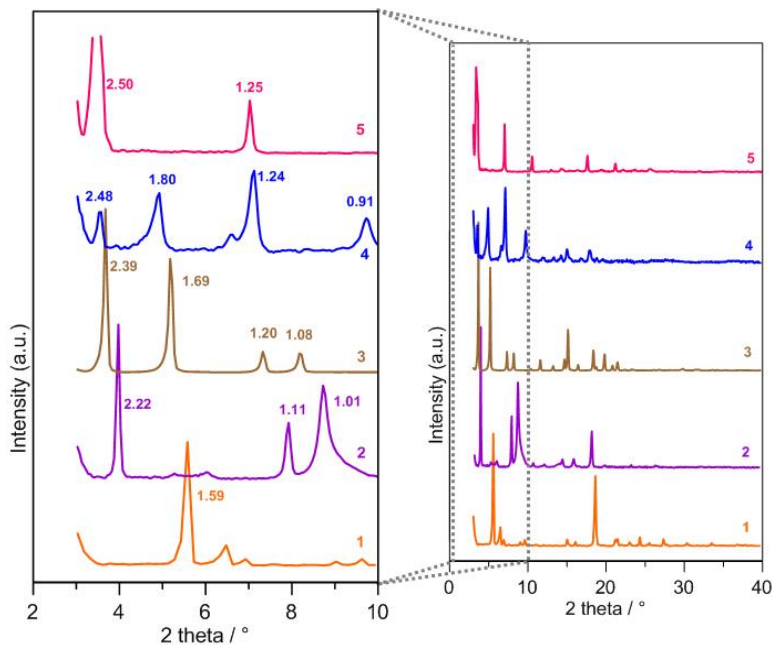

**Figure 6.** X-ray diffraction pattern of the fibers of **1-5** grown on glass.

All samples display patterns with sharp and separated peaks, flat background, presence of second order reflections, indicating highly crystalline materials. The interlayer distances of peaks registered at the lowest angular values generally give the direct measurement of translational vectors in the crystal. In this hypothesis, the fibers from **5** (*T4S4BZ*), **4** (*T4S4CF*) and **3** (*T4S4TN*), show reflections corresponding to a 2.4-2.5 nm distance, those grown from **2** (*T4S4TNO2*) show a slightly smaller distance, 2.2 nm, while the distance measured for those grown from **1** (*T4S4T*) is 1.6 nm. Putting these values together with the molecular lengths obtained from theoretical calculations, a unique packing pattern can be proposed within the unit cells having molecules differently inclined with respect to the main crystallographic axes, as sketched in Figure 7. Concerning the fibers grown from **4**, an overall decrease of the diffracted intensity and an increase of full width at half-maximum (FWHM) of the most intense peak (~0.23°) is observed, almost twice that of the other fibers (~0.11°). FWHM is a draft indicator of the crystal size: the smaller it is, the bigger the main size of the crystal domains is. This proves that in the experimental conditions used the crystals of **4** are less perfect and with shorter coherent domains than all the others, likely due to the presence of the long fluorinated terminal chains assuming disordered conformations. In Supporting Information (Figure S41) we report for comparison the X-ray diffraction patterns of the cast films of **1-5** deposited from $CH_2Cl_2$. Only the films from **1, 2** and **3** show peaks due to crystalline order, the peaks being however at different angular values with respect to those of the corresponding fibers hence indicating a different type of solid state aggregation. Moreover, worse overall quality, smaller number of peaks and bigger peak widths indicative of less order and smaller crystal domains are observed. Combining the X-ray data with the information from DFT calculations it was possible to elaborate a common model for the growth of tape-like and helical fibers. The mechanism proposed for the growth modalities of the fibers and the chirality transfer from the molecular level to the meso- and microscale of compounds **1-5** is illustrated in Figure 8.

**Figure 7.** Left: sketches of the proposed unit cells of the fibers obtained from **1-5**; axes lengths in nm are obtained from the distances of the main reflections of X-ray plots. Right: DFT calculated conformations of the dimers of **1-5**.

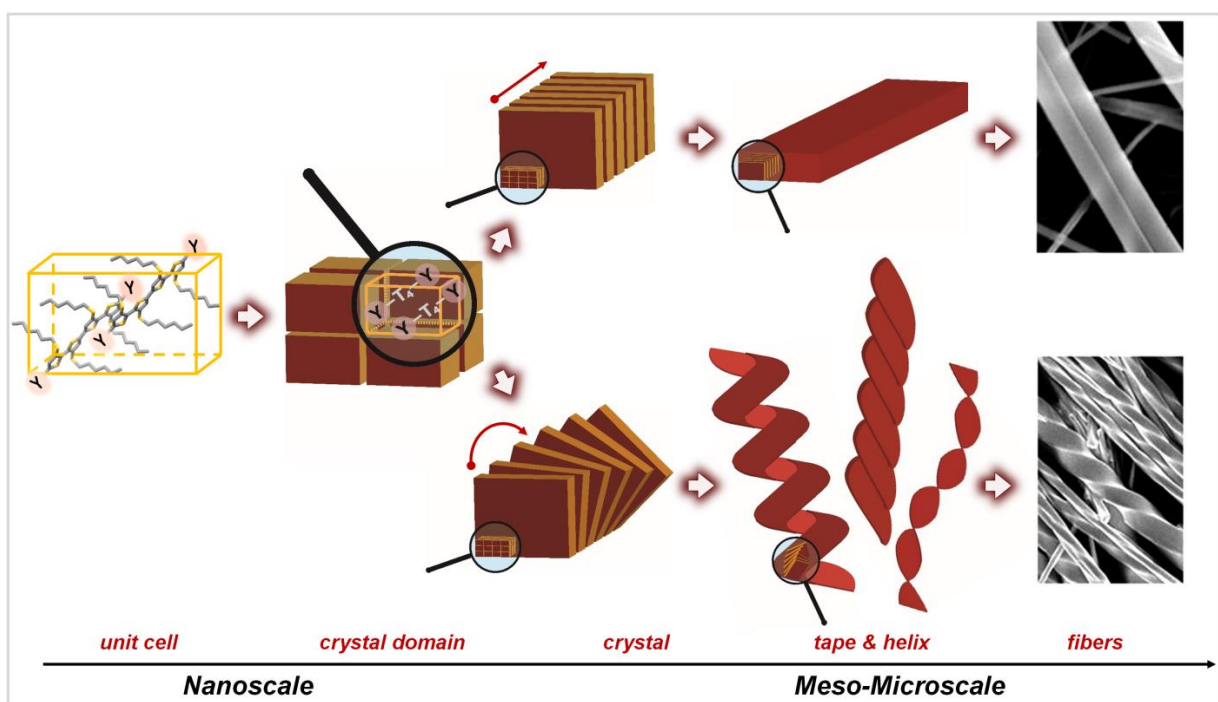

**Figure 8**. Proposed model for the formation of tape-like and helical crystalline fibers from **1-5.**

As suggested by the theoretical calculations, some torsional flexibility is present in the thiophenic backbones of all the molecules as a result of intermolecular interactions, based mainly on π−π stacking for compounds **1** and **5** and π−π stacking mediated by hydrogen bonds for **2-4**. It is reasonable to assume that this molecular torsion, which contributes to conformational chirality, is maintained inside the unit cell of the crystal. Several unit cells are coherently aligned to form the crystal domains, the linear dimension of which, as roughly estimated from the FWHM of the XRD main peak, is about 85 nm that corresponds to about 35 unit cells (35 nm for **4** corresponding to about 20 unit cells). Each crystal is formed by several domains. The crystalline unit cells assemble together to form platelets subsequently packing either as tape-like fibers as in **1-2**, or as coiled fibers as in **3-5** depending on packing forces. The chirality transfer to a larger scale takes place during the growth of the nanocrystals: regular stacking faults driven by intermolecular interactions cause the formation of elongated helices developing in the growth direction. As reported above, the long fluorurated alkyl chains of **4** have some difficulties to regularly assemble in the crystal and in consequence the crystal domains are smaller. The same is true for the crystals that enlarge twisting and shifting during the stacking process thus growing as a more screwed helix. Strong polarization and formation of important hydrogen bonds between molecules of **3** can be the reasons of a more regular interaction, resulting on the large scale in very regular helices.

*Photoconductivity.* All the fibers of **1-5** are photoconductive, i.e. they become more electrically conductive upon absorption of visible light. We evaluated the photoresponse of the fibers networks deposited on planar gold electrodes by measuring their conductivity in dark conditions and under an impinging constant illumination from a white light source with an intensity of 1 mW/cm$^2$ (Figure 9A). The individual measurements for the different fiber samples are reported in Figure S43, while Figure 9C shows the the measurements for two representative fibers, those obtained from compounds **1** and **5**. In the case of **1**, dark conductivity is very low, at the limit of our measuring setup (< 1 pA at a bias of 5 V); when irradiated, a marked photoresponse is detected, with the current increasing more than a factor of 10 (~20 pA at 5 V). In the case of **5**, which is four orders of magnitude more conductive in dark (~800 pA at 5 V), the photoresponse is much weaker, with an increase < 2. Overall, the devices drive currents in the range 1 pA to 1 nA in both conditions, at a bias of 5 V, in agreement with literature reports for oligothiophene-based self- assembled nanostructures[40] and poly(3-hexylthiophene films).[41] The morphology of the deposited active layers, being a network of semiconducting fibers, does not allow a direct comparison of the conductivities between the different samples without estimating the coverage of the fibers onto the substrate and the electrodes. We therefore restricted our analysis to the assessment of the photo-response properties of the materials through the measurement of the ratio between the conductivities in the "light" and "dark" conditions (Figure 9B).

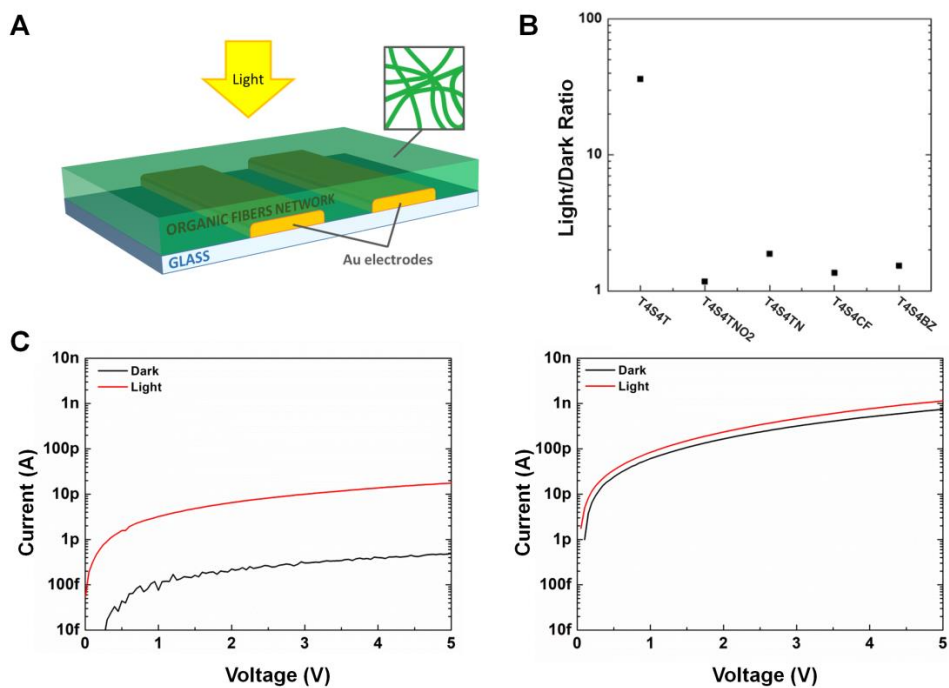

**Figure 9.** A) Device structure for the measurement of the photo-response of the fibers of **1-5** grown on photolithographically-defined gold electrodes on glass, B) Ratio of the conductivities of the materials in the "light" and "dark" conditions. C) Current-Voltage measurements for the fibers of 1 (left panel) and 5 (right panel) grown on photolithographically-defined gold electrodes on glass in the "dark" (black line) and in the "light" (red line) conditions.

Overall, four of the materials behave similarly to **5**, exhibiting a very limited light/dark ratio, lower than 2, while those obtained from **1** exhibit a much higher ratio of 36, an order of magnitude higher than the other homologue materials. While it is not possible to conclude about the photoresponse mechanism here, we observe that the highest on-off ratio is obtained in the case of the network characterized by the lowest dark current, possibly implying a photoconduction or photoinjection mechanism.[42]

***KPFM characterization.*** The of **1-5** grown on glass were analyzed by Kelvin Probe Force Microscopy (KPFM)[43] for the measurement of the surface electronic potential. In particular, the technique allows to measure the work function of the fibers with respect to the gold tip value: $\phi =$

$\phi_{fiber}$ - $\phi_{Au}$. The work functions of the fibers of **1-5** are reported in Figure 10. In agreement with previous characterizations,[44] we can clearly distinguish two classes of fibers.

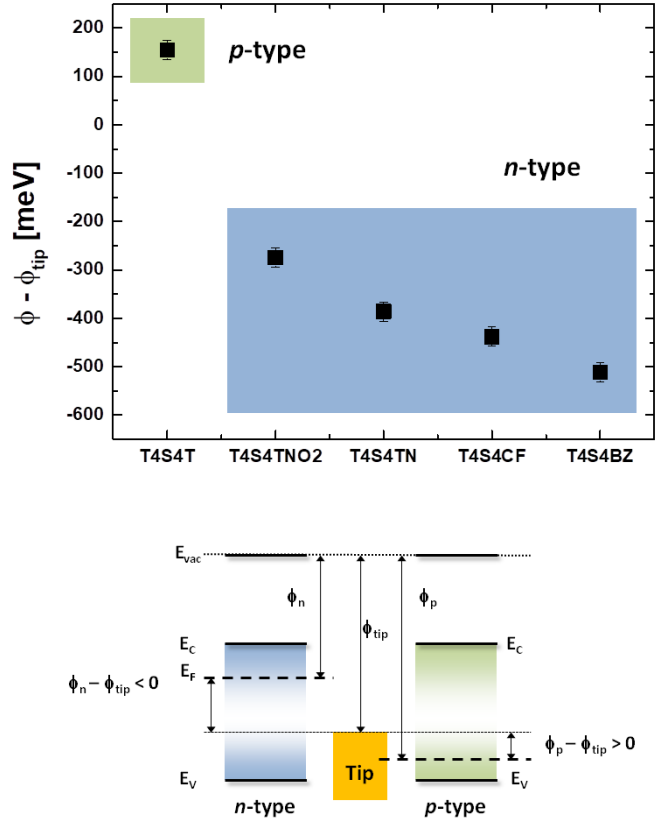

**Figure 10.** Surface electronic potential measured by KPFM for the fibers of **1-5** grown on glass and energy diagram for *p*-type and *n*-type charge carriers.

Indeed, the fibers obtained from **1** show a work function value higher than gold ($\phi - \phi_{Au} > 0$) corresponding to the increase of hole density at surface (electrons moving from the fibers to the tip) whereas the fibers obtained from **2-5** show work function values lower than gold ($\phi - \phi_{Au} < 0$) corresponding to an increase of the electron density (electrons moving from the tip to the fiber). This evidence allows to adress the fibers of **1** as a *p*-type whereas the fibers of **2-5** are *n*-type, i.e. the functionalization with electron acceptors causes the reversal of the sign of charge carriers. The results are in agreement with cyclovoltammetry data, indicating that passing from the fibers formed by **1** to those formed by **2-5** there is a marked increase in electron affinity (Table 2).

**CONCLUSIONS**

Using thiophene based oligomers we have demonstrated that it is possible to establish a straightforward correlation between the functional properties of self-assembled nano/micro- fibers and the molecular structure of the precursors. To this purpose oligomers including building blocks containing in their covalent network the information needed to promote anisotropic growth and capable to undergo modifications in their electronic distribution without altering the self-assembly modalities are required. If these conditions are fulfilled, a wide tuning of important properties is achievable. In the present work we have described one of such building blocks, i.e. a '*non covalent interactional algorithm*' (according to the definition of J. M. Lehn[11]), which enabled an unprecedented tuning of various properties within a set of comparable crystalline, electroactive, self-assembled nano/microfibers. In particular, we describe here changes in polarity of major charge carriers from *p*-type to *n*-type, light emission from visible to NIR and extended tuning of electrochemical energy gaps.

We expect that the search for novel '*non covalent interactional algorithms*' will be particularly fruitful in the field of thiophene oligomers, which are very versatile functional materials, fluorescent and conductive, with astonishing recognition capabilities *via* non-bonded interactions even inside living cells and organisms.[45] They can physiologically co-assemble with specific proteins inside live cells[15e] and are *p*-type semiconductors but can acquire *n*-type charge transport properties by sulfur functionalization with oxygen.[46] Thus, it is probably not too hazardous to believe that the identification of appropriate '*non covalent interactional algorithms*' will be beneficial not only to the search of new and more performing materials for organic electronics and optoelectronics but also to the discovery of new properties of thiophene compounds for their application in fields at the frontier of biology, medicine and supramolecular chemistry.

**ASSOCIATED CONTENT**

**Supporting Information**

Methods, synthesis of materials, $^1$H and $^{13}$C NMR Spectra, additional data of optical microscopy, emission lifetimes, cyclic voltammetry, X-ray diffraction, photoconductivity.

Supplementary information is available free of charge via the Internet at http://pubs.acs.org.

**NOTES**

The authors declare no competing financial interest.

**AUTHOR INFORMATION NOTES**

**Corresponding Author note**

giovanna.barbarella@isof.cnr.it

francesca.dimaria@isof.cnr.it

**PRESENT ADRESS NOTE**

**Present adress of RM**

Division of Materials Science, Department of Engineering Science and Mathematics

Luleå University of Technology, 971 87 Luleå, Sweden.

raffaello.mazzaro@ltu.se

**ACKNOWLEDGMENTS**

F.D.M, G.B, A.F. and D.B. acknowledge financial support from the UE project INFUSION (Engineering optoelectronic INterfaces: a global action intersecting FUndamental conceptS and technology implementatION of self-organized organic materials, Proposal number: 734834). D.G. was supported by the Italian flagship NANOMAX, N-CHEM. M. C. and A. P. acknowledges support by the European Research Council (ERC) under the European Union's Horizon 2020 research and innovation program "HEROIC," Grant Agreement No. 638059.

**REFERENCES**

1  (a) G. M. Whitesides, M. Boncheva, *PNAS* **2002**, *99*, 4769-4774. (b) B. A. Grzybowski, C. E. Wilmer, J. Kim, K. P. Browne, K. J. M. Bishop, *Soft Matter* **2009**, *5*, 1110-1128. (c) B. A. Grzybowski, K. Fitzner, J. Paczesny, S. Granick, *Chem. Soc. Rev.* **2017**, *46*, 5647-5678. (d) *Molecular Self-assembly in Nanoscience and Nanotechnology*, A. Kilislioğlu and S. Karakus, Eds. InTech, **2017**, DOI: 10.5772/65607.

2  Y. Guo, L. Xu, H. Liu, Y. Li, C. M. Che, Y. Li, *Adv. Mater.* **2015**, *27*, 985-1013.

3  Y. Gong, Q. Hu, N. Cheng, Y. Bi, W. Xuac, L. Yu, *RSC Adv.* **2015**, *5*, 32435-32440.

4	F. S. Kim, G. Ren, S. A. Jenekhe, *Chem. Mater.* **2011**, *23*, 682-732.

5	Y. Yamamoto, *Sci. Technol. Adv. Mater.* **2012**, *13*, 033001 (15pp).

6	S. Chen, P. Slattum, C. Wang, L. Zang, *Chem. Rev.* **2015**, *115*, 11967-11998

7	E. Moulin, J. J. Cid, N. Giuseppone, *Adv. Mater.* **2013**, *25*, 477-487.

8	E. Yashima, N. Ousaka, D. Taura, K. Shimomura, T. Ikai, K. Maeda, *Chem. Rev.* **2016**, *116*, 13752-13990.

9	S. I. Stupp, L. C. Palmer *Chem. Mater.* **2014**, *26*, 507-518.

10	T. Aida, E.W. Meijer, S. I. Stupp, *Science* **2012**, *335*, 813-817.

11	J. M. Lehn, *Angew. Chem. Int. Ed.* **2015**, *54*, 3276-3289.

12	L. Zhang, N. S. Colella, B. P. Cherniawski, S. C. B. Mannsfeld, A. L. Briseno, *ACS Appl. Mater. Interfaces* **2014**, *6*, 5327-5343.

13	D. H. Kim, J. T. Han, Y. D. Park, Y. Jang, J. H. Cho, M. Hwang, K. Cho, *Adv. Mater.* **2006**, *18*, 719-723.

14	D. A. Stone, A. S. Tayi, J. E. Goldberger, L. C. Palmer, S. I. Stupp, *Chem.Commun.* **2011**, *47*, 5702-5704.

15	(a) F. Di Maria, P. Olivelli, M. Gazzano, A. Zanelli, M. Biasiucci, G. Gigli, D.Gentili, P. D'Angelo, M. Cavallini, G. Barbarella, *J. Am. Chem. Soc.* **2011**, *133*, 8654-8661. (b) F. Di Maria, E. Fabiano, D. Gentili, M. Biasiucci, T. Salzillo, G. Bergamini, M. Gazzano, A. Zanelli, A. Brillante, M. Cavallini, F. Della Sala, G. Gigli, G. Barbarella, *Adv. Funct. Mater.* **2014**, *24*, 4943-4951. (c) F. Di Maria, M. Zangoli, G. Barbarella, *Advances in Heterocyclic Chemistry* **2017**, *123*, 105-167. d) D. Gentili, F. Di Maria, F. Liscio, L. Ferlauto, F. Leonardi, L. Maini, M. Gazzano, S. Milita, G. Barbarella, M. Cavallini, *J. Mater. Chem.* **2012**, *22*, 20852-20856.

16	D. H. Kim, D. Y. Lee, H. S Lee, W. H. Lee, Y. H. Kim, J. I. Han, K. Cho, *Adv. Mater.* **2007**, *19*, 678-682.

17	G. Pescitelli, L. Di Bari, N. Berova, *Chem. Soc. Rev.* **2014**, *43*, 5211-5233.

18	S. Arias, R. Rodríguez, E. Quiñoá, R. Riguera, F. Freire, *J. Am. Chem. Soc.* **2018**, *140*, 667-674.

19	M. D. Shoulders, R. T. Raines, *Annu. Rev. Biochem.* **2009**, *78*, 929-958.

20	M. Hifsudheen, R. K. Mishra, B. Vedhanarayanan, V. K. Praveen, A. Ajayaghosh *Angew. Chem. Int. Ed.* **2017**, *56*, 1-6.

21  M. Cavallini, D. Gentili, P. Greco, F. Valle, F. Biscarini, *Nat. Protoc.* **2012**, *7*, 1668-1676.

22  G. Crosby, J. Demas, *J. Phys. Chem.* **1971**, *75* (8), 991-1024.

23  K. Suzuki, A. Kobayashi, S. Kaneko, K. Takehira, T. Yoshihara, H. Ishida, Y. Shiina, S. Oishi, S. Tobita, *Phys. Chem. Chem. Phys.* **2009**, *11* (42), 9850-9860.

24  J. C. De Mello, H. F. Wittmann, R. H. Friend, *Adv. Mater.* **1997**, *9* (3), 230-232.

25  H. Chosrovian, S. Rentsch, D. Grebner, D. U. Dahm, E. Birckner, H. Naarmann, *Synth. Met.* **1993**, *60* (1), 23-26.

26  M. Theander, O. Inganäs, W. Mammo, T. Olinga, M. Svensson, M. R. Andersson, *J. Phys. Chem. B* **1999**, *103* (37), 7771-7780.

27  A. Yang, M. Kuroda, Y. Shiraishi, T. Kobayashi, *J. Chem. Phys.* **1998**, *109* (19), 8442-8450.

28  L. Antolini, E. Tedesco, G. Barbarella, L. Favaretto, G. Sotgiu, M. Zambianchi, D. Casarini, G. Gigli, R. Cingolani, *J. Am. Chem. Soc.* **2000**, *122* (37), 9006-9013.

29  TURBOMOLE, V7.1; TURBOMOLE GmbH: Karlsruhe, Germany, **2011**. http://www.turbomole.com.

30  F. Furche, R. Ahlrichs, C. Hättig, W. Klopper, M. Sierka, F. Weigend, *Wiley Interdiscip. Rev. Comput. Mol. Sci.* **2014**, *4*, 91-100.

31  L. A. Constantin, E. Fabiano, F. Della Sala, *J. Chem. Theory Comput.* **2013**, *9*, 2256-2263.

32  L. A. Constantin, E. Fabiano, F. Della Sala, *Phys.Rev.B* **2012**, *86*, 035130.

33  L. A. Constantin, E. Fabiano, F. Della Sala, *Phys. Rev. B* **2013**, *88*, 125112.

34  M. Sierka, A. Hogekamp, R. Ahlrichs, *J. Chem. Phys.* **2003**, *118*, 9136.

35  F. Weigend, M. Haser, H. Patzelt, R. Ahlrichs, *Chem. Phys. Lett.* **1998**, *294*, 143-152.

36  A. Schäfer, H. Horn, R. Ahlrichs, *J. Chem. Phys*. **1992**, *97*, 2571.

37  E. R. Johnson, S. Keinan, P. Mori-Sanchez, J. Contreras-Garcia, A. J. Cohen, W.Yang, *J. Am. Chem. Soc.* **2010**, *132*, 6498-6506.

38  J. Contreras-Garcia, W. Yang, E. R. Johnson, *J. Phys. Chem. A* **2011**, *115*, 12983-12990.

39  J. Contreras-Garcia, E. R. Johnson, S. Keinan, R. Chaudret, J. P. Piquemal, D. N. Beratan, W. Yang, *J. Chem. Theory Comput.* **2011**, *7*, 625-632.

40  D. A. Stone, A. S. Tayi, J. E. Goldberger, L. C. Palmera, S. I. Stupp, *Chem. Commun.* **2011**, *47*, 5702-5704.

41  N. V. Tkachenko, V. Chukharev, P. Kaplas, A. Tolkki, A. Efimov, K. Haring, J. Viheria, T. N. H. Lemmetyinen, *Appl. Surf. Sci.* **2010**, *256*, 3900-3905.

42  K-J; Baeg, M. Binda, D. Natali, M. Caironi, Y. Y. Noh, *Adv. Mater.* **2013**, *25*, 4267-4295.

43  (a) A. Liscio, V. Palermo, P. Samorí, *Acc. Chem. Res.* **2010**, *43*, 541-550.
(b) C. Musumeci, A. Liscio, V. Palermo, P. Samorì, *Materials Today* **2014**, *17*, 504-517.

44  F. Di Maria, A. Zanelli, A. Liscio, A. Kovtun, E. Salatelli, R. Mazzaro, V. Morandi, G. Bergamini, A. Shaffer, S. Rozen, *ACS Nano* **2017**, *11*, 1991-1999.

45  K. M. Psonka-Antonczyk, P. Hammarström, L. B. G. Johansson, M. Lindgren, B. T. Stokke, K. P. R. Nilsson, S. Nyström, *Front. Chem.* **2016**, *4*, 44.

46  (a) J. Z. Low, B. Capozzi, J. Cui, S. Wei, L. Venkataraman, L. M. Campos, *Chem. Sci.* **2017**, 3254-3259. (b) B. Capozzi, J. Xia, O. Adak, E. J. Dell, Z.F. Liu, J. C. Taylor, J. B. Neaton, L. M. Campos, L. Venkataraman, *Nat. Nanotechnol*. **2015**, *10*, 522-527.